\documentclass
[superscriptaddress,secnumarabic,amssymb,amsmath,nobibnotes,aps,prd,showkeys,showpacs,nofootinbib,onecolumn]{revtex4}%
\usepackage{graphicx}
\usepackage{epsf}
\usepackage{bm}
\usepackage{amsmath}
\usepackage{amsfonts}
\usepackage{amssymb}
\usepackage{epstopdf}
\usepackage{mwe}
\usepackage{caption}%
\providecommand{\U}[1]{\protect\rule{.1in}{.1in}}

\newcommand{\be}{\begin{equation}}
\newcommand{\ee}{\end{equation}}

\newcommand{\mincir}{\raise
-3.truept\hbox{\rlap{\hbox{$\sim$}}\raise4.truept\hbox{$<$}\ }}
\newcommand{\magcir}{\raise
-3.truept\hbox{\rlap{\hbox{$\sim$}}\raise4.truept\hbox{$>$}\ }}

\ifx\pdfoutput\relax\let\pdfoutput=\undefined\fi
\newcount\msipdfoutput
\ifx\pdfoutput\undefined\else
\ifcase\pdfoutput\else
\ifx\paperwidth\undefined\else
\ifdim\paperheight=0pt\relax\else\pdfpageheight\paperheight\fi
\ifdim\paperwidth=0pt\relax\else\pdfpagewidth\paperwidth\fi
\fi\fi\fi
\begin{document}
\title{Noether symmetries and stability of ideal gas solution in Galileon Cosmology}
\author{N. Dimakis}
\email{nsdimakis@gmail.com}
\affiliation{Instituto de Ciencias F\'{\i}sicas y Matem\'{a}ticas, Universidad Austral de
Chile, Valdivia, Chile}
\author{Alex Giacomini}
\email{alexgiacomini@uach.cl}
\affiliation{Instituto de Ciencias F\'{\i}sicas y Matem\'{a}ticas, Universidad Austral de
Chile, Valdivia, Chile}
\author{Sameerah Jamal}
\email{sameerah.jamal@wits.ac.za}
\affiliation{School of Mathematics and Centre for Differential Equations,Continuum
Mechanics and Applications, University of the Witwatersrand, Johannesburg,
South Africa}
\author{Genly Leon}
\email{genly.leon@pucv.cl}
\affiliation{Instituto de F\'{\i}sica, Pontificia Universidad Cat\'olica de
Valpara\'{\i}so, Casilla 4950, Valpara\'{\i}so, Chile}
\author{Andronikos Paliathanasis}
\email{anpaliat@phys.uoa.gr}
\affiliation{Instituto de Ciencias F\'{\i}sicas y Matem\'{a}ticas, Universidad Austral de
Chile, Valdivia, Chile}
\affiliation{Institute of Systems Science, Durban University of Technology, PO Box 1334,
Durban 4000, Republic of South Africa}

\begin{abstract}
A class of generalized Galileon cosmological models, which can be described by
a point-like Lagrangian, is considered in order to utilize Noether's Theorem
to determine conservation laws for the field equations. In the
Friedmann-Lema\^{\i}tre-Robertson-Walker universe, the existence of a
nontrivial conservation law indicates the integrability of the field
equations. Due to the complexity of the latter, we apply the differential
invariants approach in order to construct special power-law solutions and
study their stability.

\end{abstract}
\maketitle
\date{\today}

\section{Introduction}

\label{sec:level1}

The accuracy of the new cosmological observations
\cite{Teg,Kowal,Komatsu,Ade15,planck2015} leads us to the necessity of
extending the General theory of Relativity. The introduction of a scalar
field, which attributes the effects of the so-called dark energy, in the
Einstein-Hilbert action is proposed in order to provide new mechanisms in
order to explain the various phases of the universe
\cite{Brans,ref1a,ref1b,Hanlon,ratra,peebles,pha,tsujikawa1}. Canonical scalar
fields lead to second-order gravitational theories while at the same time,
they can describe the dynamics of higher-order theories of gravity or
higher-order terms of non-canonical fields, see \cite{s01,s02,s03} and
references therein. However, it is possible for a non-canonical scalar field
Lagrangian to provide up to second-order system of differential equations when
the Lagrangian is that of Horndeski theories \cite{hor}. A special family of
non-canonical fields which have been proposed are the so-called Galileons
\cite{nik,gal02}. The cosmological theory which include the action integral of
the Galileons is called Galileon cosmology - various cosmological applications
of the theory can be found in
\cite{m01,m02,m03,m04,m05,m06,m07,m08,m09,m10,m11a,m11,m12}. However, while
Galileon cosmology is a second-order gravitational theory and has been well
studied in the literature, the existence of actual solutions for the field
equations is still an open subject, and it is the reason that motivates this work.

Indeed, in Galileon cosmology the gravitational field equations are at most of
second-order while in an isotropic and homogeneous universe, that is in a
Friedmann-Lema\^{\i}tre-Robertson-Walker (FLRW) background geometry, there are
two degrees of freedom which describe the scale factor $a\left(  t\right)  $
and the Galileon field $\phi\left(  t\right)  $. The derivation of a solution
for a system of differential equations that can be expressed in terms of
elementary functions is usually linked to the existence of a sufficient number
of independent first integrals, that is conservation laws; or equivalently to
the existence of a sufficient number of transformations which reduce the
differential equations to a system of algebraic equations. When this is true
we say that the dynamical system is integrable. These properties are related
and in a large extent equivalent to the existence of symmetries. In this work
and in order to study the integrability of the Galileon cosmological model
under consideration, we search for conservation laws of the action.

We find that the gravitational field equations can follow from the variation
of a point-like Lagrangian and by applying Noether's first theorem we
construct the conservation laws of the system. The class of transformations we
consider to leave invariant the action and, hence, the field equations are the
so-called point transformations. The latter have been applied in various
cosmological models and modified theories of gravity for the determination of
new cosmological solutions, see for instance
\cite{ref1,ref2,ref3,ref4,ref5,ref6,ref7,ref8,basil,ngp} and references
therein. It is important to mention that except for the point transformations,
other classes of mappings or different methods for the study of integrability
have been applied in gravitational theories, some of these are presented in
\cite{Aref1,Aref2,Aref3,Aref4,Aref5,Aref6,Aref7,Aref8}. Furthermore, because
the spatial field equations are differential equations of second-order bound
by a constraint, while the degrees of freedom are two, we need only determine
the unknown functions/parameters of the model whereby the system admits a
single nontrivial Noether symmetry, so as to yield it fully integrable.
Finally, in order to study the physical properties of the models that we
determine from the application of Noether's theorem, we perform a study of the
stability of powerlaw solutions.

The plan of the paper is the following: In Section \ref{section2} we present
the gravitational field equations for the model that we consider in this work.
Specifically for the Galileon's Lagrangian, we select the functions in order
to describe the so-called cubic Galileon field. The field equations of that
model have cubic power of the first derivative of the field $\phi\left(
t\right)  $, while there are two unknown functions: $V\left(  \phi\right)  $
which corresponds to the potential that forces the field and $g\left(
\phi\right)  $, which couples the higher-power terms of the field equations.
However, in the limit where $g\left(  \phi\right)  \rightarrow0$, the Galileon
field reduces to that of a scalar field minimally coupled to gravity. The
application of Noether's theorem for the model of our consideration is given
in Section \ref{section3}, while power-law solutions (ideal gas solutions) are
presented as special solutions for the field equations in Section
\ref{section4}. Moreover, the stability of those special solutions is studied.
Finally we discuss our results in Section \ref{conc}.

\section{Galileon Cosmology}

\label{section2}

The cosmological models with Galileon fields belong to the four-dimensional
scalar-tensor theories in which the spatial gravitational field equations are
of second-order and consequently are free of Ostrogradski instability
\cite{Ostrogradsky:1850fid}.

The Lagrangian of the generalized Galileon field is given to be
\cite{DeFelice:2011bh}
\begin{align}
\mathcal{L}_{G}=K-G_{3}\Box\phi+G_{4}R+G_{4,X}[(\Box\phi)^{2}-(\nabla_{\mu
}\nabla_{\nu}\phi)(\nabla^{\mu}\nabla^{\nu}\phi)]  & \nonumber\\
+G_{5}G_{\mu\nu}(\nabla^{\mu}\nabla^{\nu}\phi)-\frac{1}{6}G_{5,X}[(\Box
\phi)^{3}-3(\Box\phi)(\nabla_{\mu}\nabla_{\nu}\phi)(\nabla^{\mu}\nabla^{\nu
}\phi)  & \nonumber\\
+2(\nabla^{\mu}\nabla_{\alpha}\phi)(\nabla^{\alpha}\nabla_{\beta}\phi
)(\nabla^{\beta}\nabla_{\mu}\phi)],  &  \label{lan.01}%
\end{align}
in which the functions $K$ and $G_{i}$ ($i=3,4,5$) depend on the scalar field
$\phi$ and its kinetic energy $X=-\partial^{\mu}\phi\partial_{\mu}\phi/2$,
while $R$ is the Ricci scalar, and $G_{\mu\nu}$ is the Einstein tensor. The
above Lagrangian can be seen to be equivalent to that of the Horndeski theory
\cite{Kobayashi:2011nu}.

Among the infinite number of models which can be derived from (\ref{lan.01}),
let us assume the special case which leads to a generalized Galileon model
with cubic derivative interaction term, previously studied in \cite{genlyGL}
for the matter case, and in \cite{genlyGL2} for the Galileon vacuum, where the
free functions are
\begin{equation}
K=X-V\left(  \phi\right)  ~,~G_{3}=-g(\phi)X~,~G_{4}=\frac{1}{2},~G_{5}=0 .
\label{lan.02}%
\end{equation}
Then the gravitational Action Integral of Galileon cosmology
\begin{equation}
S=\int d^{4}x\sqrt{-g}\left(  \frac{1}{2}R-\frac{1}{2}\partial^{\mu}%
\phi\partial_{\mu}\phi-V(\phi)-\frac{1}{2}g(\phi)\partial^{\mu}\phi
\partial_{\mu}\phi\Box\phi\right)  +\int d^{4}x\sqrt{-g}L_{m} \label{lan.03}%
\end{equation}
can be written in the form%
\begin{equation}
S=S_{GR}+S_{SF}+S_{EGT}+S_{m}, \label{lan.04}%
\end{equation}
where $L_{m}$ denotes the Lagrangian of the matter source, $S_{GR}$ is the
Einstein-Hilbert Action-Integral, $S_{SF}$ \ is the action term which
corresponds to a minimally coupled scalar field and $S_{EGT}$ is the new term
which is introduced by the Galileon and includes the cubic derivative term.

Indeed, other definitions of the free functions of (\ref{lan.01}) exist and
provide us with different types of models. However, the selection of $K$ and
$G_{3}$ to be linear on $X$ is the first extension of scalar-field cosmology,
and actually in the limit $g\left(  \phi\right)  \rightarrow0$, (canonical)
scalar field cosmology is recovered.

\subsection{FLRW Cosmology}

We assume the cosmological scenario that our universe is described by the
spatially-flat FLRW spacetime metric with line element%
\begin{equation}
ds^{2}=-dt^{2}+a^{2}\left(  t\right)  \left(  dx^{2}+dy^{2}+dz^{2}\right)  ,
\label{bd.05}%
\end{equation}
for which the Ricci scalar is\
\begin{equation}
R=6\left[  \frac{\ddot{a}}{a}+\left(  \frac{\dot{a}}{a}\right)  ^{2}\right]  .
\label{bd.06}%
\end{equation}

Consequently, if we assume that the isometries of (\ref{bd.05}) are also
inherited by the matter fields, then it follows that $\phi\equiv\phi\left(
t\right)  $. This means that the Galileon field possesses the symmetries of
the spacetime and the gravitational field equations are ordinary differential
equations (ODEs). Let us present the gravitational field equations of the
action (\ref{lan.03}).

Variation with respect to the metric tensor and the scalar field of the action
integral (\ref{lan.03}) provides the Friedmann equation
\begin{equation}
3H^{2}=\frac{\dot{\phi}^{2}}{2}\left(  \,1-6g(\phi)H\dot{\phi}+g^{\prime}%
(\phi)\,\dot{\phi}^{2}\right)  +V(\phi), \label{ss1}%
\end{equation}
the acceleration equation
\begin{equation}
2\dot H+\dot\phi^{2}\left(  1+ g^{\prime}(\phi)\,\dot\phi^{2}-3 g(\phi
)\,H\,\dot\phi+g(\phi)\,\ddot\phi\right)  =0 \label{f1}%
\end{equation}
and the Klein-Gordon-like equation
\begin{equation}
\ddot\phi\left(  2 \dot\phi^{2} \, g^{\prime}(\phi)-6 H \, g(\phi)\, \dot
\phi+1\right)  +\dot\phi^{2} \left(  \frac{1}{2} \dot\phi^{2} \,
g^{\prime\prime}(\phi)-3 \, g(\phi)\, \dot H-9 \, H^{2} \, g(\phi)\right)  +3
\, H \, \dot\phi+V^{\prime}(\phi)=0 \label{f2}%
\end{equation}
where we have assumed that there is no any extra matter source.

A substitution of (\ref{bd.06}) into (\ref{lan.03}) followed by integration by
parts derives the Lagrange function
\begin{equation}
\mathcal{L}\left(  a,\dot{a},\phi,\dot{\phi}\right)  = 3\,a\, \dot{a}%
^{2}-\frac{1}{2}\,{a}^{3}\dot{\phi}^{2} +{a}^{3}V(\phi)+g(\phi)a^{2}\dot{a}\,
\dot{\phi}^{3}-\frac{g^{\prime}(\phi)}{6}{a}^{3}\,\dot{\phi}^{4}, \label{lag}%
\end{equation}
from which the two spatial gravitational field equations (\ref{f1}) and
(\ref{f2}) can be derived with the action of the Euler operator. Here, it is
important to remark that the first Friedmann equation is a constraint equation
and can be derived from (\ref{lag}) if we re-instate the gauge invariance in
(\ref{bd.05}) by introducing an arbitrary lapse function $N\left(  t\right)
$. At this point, a comment is in order. Although the gauge fixing process is
trivial at the level of the equations of motion, this is not true when it
takes place in the Lagrangian \cite{Japonezoi}. That is why, by considering
(\ref{lag}), we have to additionally simulate the constraint equation of
motion with a first integral which necessarily needs to be zero. The existence
of such a conserved quantity is guaranteed due to fact that the system is
autonomous. We also have to keep in mind that the non-gauged fixed version of
a system of this type admits in general different groups of symmetries
\cite{tchris}. However, for the model under consideration, it can be easily
checked that the exact same conserved quantity emerges through Noether's first
theorem in both considerations. In that respect, we can proceed by fixing
$N=1$ at the Lagrangian level.

As we did for the action before, we can write Lagrangian (\ref{lag}) in the
form%
\begin{equation}
\mathcal{L}\left(  a,\dot{a},\phi,\dot{\phi}\right)  =\mathcal{L}%
_{GR}+\mathcal{L}_{SF}+\mathcal{L}_{EGT} \label{lag.01}%
\end{equation}
where%
\begin{equation}
\mathcal{L}_{EGT}=g(\phi)a^{2}\dot{a}\, \dot{\phi}^{3}-\frac{g^{\prime}(\phi
)}{6}{a}^{3}\,\dot{\phi}^{4} \label{lag.02}%
\end{equation}
and the rest of the terms of (\ref{lag.01}) give the Lagrangian of a minimally
coupled scalar field in a spatially flat FLRW spacetime. Function (\ref{lag})
is a point-like Lagrangian, however it differs from that of scalar tensor
theories because of the term $\mathcal{L}_{EGT}$ which introduces a cubic
first-derivative dependent force in the evolution of motion. Therefore the
momenta are calculated to be%
\begin{equation}
p_{a}=6\, a\dot{a}+g\left(  \phi\right)  a^{2}\dot{\phi}^{3},
\end{equation}%
\begin{equation}
p_{\phi}=\frac{a^{2}\dot{\phi}}{3}\left(  9 g\left(  \phi\right)  \dot{a}\,
\dot{\phi}-2 a g^{\prime}\left(  \phi\right)  \dot{\phi}^{2}-3 a\right)  .
\end{equation}

We continue our analysis with the determination of the unknown functions
$g\left(  \phi\right)  $ and $V\left(  \phi\right)  $, in which the Lagrangian
(\ref{lag}) admits Noether (point) symmetries and the corresponding group of
transformations under which the action remains form invariant.

\section{Point transformations and Noether symmetries}

\label{section3}

In this work we are interested in point transformations, and for convenience
of the reader we discuss some preliminary material important for the analysis
which follows. Consider a system of second-order ordinary differential
equations
\begin{equation}
\ddot{x}^{i}=\omega^{i}\left(  t,x^{j},\dot{x}^{j}\right)  . \label{Lie.0}%
\end{equation}
where $t$ is the independent variable and $x^{i}$ denotes the dependent variables.

An one-parameter point transformation in the space $\left\{  t,x^{j}\right\}
$, such that $\left\{  t,x^{j}\right\}  \rightarrow\left\{  \bar{t}\left(
t,x^{j},\varepsilon\right)  ,\bar{x}^{j^{\prime}}\left(  t,x^{j}%
,\varepsilon\right)  \right\}  $ has the property of mapping solutions of
(\ref{Lie.0}) to themselves, satisfies in infinitesimal form the following
criterion of invariance
\[
X^{\left[  2\right]  }\left(  \ddot{x}^{i}-\omega^{i}\right)  =0
\quad\mathrm{mod} \quad\ddot{x}^{i}-\omega^{i}=0,
\]
where its generator $X$ is defined as
\begin{equation}
X=\frac{\partial\bar{t}}{\partial\varepsilon}\Bigg|_{\varepsilon=0}%
\partial_{t}+\frac{\partial\bar{x}^{i}}{\partial\varepsilon}%
\Bigg|_{\varepsilon=0}\partial_{i} \label{Lie.01}%
\end{equation}
and $X^{\left[  N\right]  }$ indicates the $N$th extension of $X$ in the jet
space $\left\{  t,x^{j},x^{\left(  1\right)  j},...,x^{\left(  N\right)
j}\right\}  ~$\cite{StephaniB}. The generator of the point transformation $X$
is called a\ Lie symmetry for the system of differential equations. Symmetries
are important in any physical system and play a significant role in every
physical theory, especially after the famous work of Emmy Noether published in
1918 \cite{Noetherpaper}.

What is more, let us consider that the system (\ref{Lie.0}) follows from the
variation of the action integral $S=\int\mathcal{L}dt.$ The first of Noether's
theorems states that when a (finite) group of transformations leaves the
action form invariant, i.e. when (for the mono-parametric transformation
considered here)
\begin{equation}
\label{Noethcon}S\left(  t,x^{j},...\right)  = S\left(  \bar{t}\left(
t,x^{j},\varepsilon\right)  ,\bar{x}^{j}\left(  t,x^{j},\varepsilon\right)
,...\right)  ,
\end{equation}
then a conserved quantity exists which can be constructed with the help of the
symmetry generator. In the case of a Lagrangian containing up to first order
derivatives of the $x^{j}$'s like in our case, condition \eqref{Noethcon} is
expressed in infinitesimal form as
\begin{equation}
X^{\left[  1\right]  }\mathcal{L}+\mathcal{L}\frac{d}{dt}\left(
\frac{\partial\bar{t}}{\partial\varepsilon}\Bigg|_{\varepsilon=0}\right)
=\dot{f} \label{Lie.03}%
\end{equation}
and the relative integral of motion is given by
\begin{equation}
I= \left(  \dot{x}^{j}\frac{\partial\mathcal{L}}{\partial\dot{x}^{j}%
}-\mathcal{L}\right)  \frac{\partial\bar{t}}{\partial\varepsilon
}\Bigg|_{\varepsilon=0} -\frac{\partial\mathcal{L}}{\partial\dot{x}^{j}}
\frac{\partial\bar{x}^{j}}{\partial\varepsilon}\Bigg|_{\varepsilon=0} +f.
\end{equation}
The generator $X$ is called a Noether symmetry and it is straightforward to
show that it is also a Lie symmetry of the equations of motion; however the
inverse is not always true.

Of course conservation laws can be derived without the use of Noether's
theorems by using other methods. Nevertheless, the simplicity and the
globalization of the applications of Noether's work make them unique in all
areas of science and not only in physics.

From the Lagrangian (\ref{lag.01}) of the Galileon cosmological models we read
that the dependent variables are the scale factor $a\left(  t\right)  $ and
the field $\phi\left(  t\right)  $, in which $t$ is the independent variable.
Hence we assume the generator of the infinitesimal transformation to be
\begin{equation}
X=\xi\left(  t,a,\phi\right)  \partial_{t}+\eta_{a}\left(  t,a,\phi\right)
\partial_{a}+\eta_{\phi}\left(  t,a,\phi\right)  \partial_{\phi}.
\end{equation}

Moreover, as we discussed, we have to consider the first Friedmann equation
i.e. the constraint, as a zero valued integral of motion $\left(  \dot{x}%
^{j}\frac{\partial\mathcal{L}}{\partial\dot{x}^{j}}-\mathcal{L}\right)  =0$,
that is, the general form of the Noetherian conservation law for our problem
reads
\begin{equation}
I=f-\eta_{a}\frac{\partial\mathcal{L}}{\partial\dot{a}}-\eta_{\phi}%
\frac{\partial\mathcal{L}}{\partial\dot{\phi}}.
\end{equation}

We continue with the determination of the unknown functions of Lagrangian
(\ref{lag.01}) in which the field equations are invariant under the action of
point transformations.

\subsection{Symmetry analysis}

From the symmetry condition (\ref{Lie.03}) for the Lagrangian (\ref{lag.01})
of the minimally coupled Galileon field we derive the condition
\begin{equation}
\left(  X^{\left[  1\right]  }+\dot{\xi}\right)  \left(  \mathcal{L}%
_{GR}+\mathcal{L}_{SF}\right)  +\left(  X^{\left[  1\right]  }+\dot{\xi
}\right)  \mathcal{L}_{EGT}=\dot{f} \label{lie03a}%
\end{equation}
from where we define a set of constraint equations in order for the
coefficient terms of the derivatives of $\dot{a}$ and $\dot{\phi}$ in
(\ref{lie03a}) to vanish. The ensuing system is presented in Appendix
\ref{appenA}. However since $\mathcal{L}_{EGT}$ has higher-polynomial
derivatives from the other two terms of the Lagrangian (\ref{lag.01}), it is
expected that the potential which is to be defined, should assume the form of
the one in which a minimally scalar field admits an extra Noether symmetry. In
the following we assume that $g\left(  \phi\right)  $ is not constant or zero.
It is important to mentioned that the analysis we perform here it is different
from the one which presented in \cite{ngp}. The main reason is that in this
work we see the Lagrangian as a regular system in contrary to \cite{ngp} which
we saw it as a singular system. The two different approaches are
complementary, for details see the discussion in \cite{nss01,tchris}.

As we discussed above the field equations constitute an autonomous system and
admit the Noether symmetry $\partial_{t}$ for arbitrary functions $V\left(
\phi\right)  $ and $g\left(  \phi\right)  $. However in the specific case in
which%
\begin{equation}
V(\phi)=V_{0}e^{-\lambda\phi}\quad\text{and}\quad g(\phi)=g_{0}e^{\lambda\phi
}. \label{pot}%
\end{equation}
an extra Noetherian symmetry exists, viz.
\begin{equation}
X=t\partial_{t}+\frac{a}{3}\partial_{a}+\frac{2}{\lambda}\partial_{\phi},
\label{ns}%
\end{equation}
with the corresponding conservation law
\begin{equation}
I_{1}=-\left(  {2\,a^{2} \dot{a}{{-}}\frac{{{2}}}{\lambda}a^{3}\dot{\phi}%
}\right)  {+g_{0}{\mathrm{e}^{\lambda\phi}}a^{3}\dot{\phi}^{3}-}\frac
{6}{\lambda} g_{0} a^{2}e^{\lambda\phi}\dot{a}\dot{\phi}^{2}. \label{con01}%
\end{equation}

Recall that the same conservation law exists in the limit in which $V_{0}=0$.
An important observation is that when the universe is dominated by the
potential of the scalar field, then \thinspace$g\left(  \phi\right)
\rightarrow0$, and the model reduces to that of a minimally coupled scalar field.

As we can see the linear term of the conservation law is that of a minimally
coupled scalar field while the nonlinear terms follow from the $\mathcal{L}%
_{EGT}$ of the gravitational Lagrangian. Therefore with the use of the
conservation law (\ref{con01}) and the constraint equation (\ref{ss1}) the
gravitational field equations are reduced to a system of two-first order ODEs
which are autonomous.

\section{Ideal gas solution as group invariant solution}

\label{section4}

Above, we made it clear that a Noether symmetry is always a Lie symmetry,
where the latter means that there exists a set of invariants in which the
equations are independent. However it is easier to construct invariant solutions.

In this respect, for the gravitational field equations (\ref{ss1})-(\ref{f2})
we find the following set of solutions:

The power-law solution
\begin{equation}
\label{solution1}a_{1}\left(  t\right)  =a_{0}t^{p}~,~\phi\left(  t\right)
=\frac{2}{\lambda}\ln\left(  \phi_{0}t\right)  ~,~g_{0}=\frac{\lambda\left(
2- \lambda^{2}p\right)  }{4(3p-1)\phi_{0}^{2}},~V_{0}=\phi_{0}^{2}\left(
\frac{2}{\lambda^{2}}+p(3p-2)\right)
\end{equation}
and the cubic root solutions
\begin{equation}
a_{2,3}\left(  t\right)  =a_{0}t^{\frac{1}{3}}~,~\phi_{2,3}\left(  t\right)
=\pm\frac{\sqrt{6}}{3}\ln(\phi_{0}t)~,~V_{0}=0~,~\lambda_{2,3}=\pm\sqrt{6},
\end{equation}
The perfect fluid solution $a_{1}\left(  t\right)  $ is similar to the special
solution for the exponential potential of a minimally coupled scalar field. We
mentioned that the terms $\mathcal{L}_{EGT}$ in the first Friedmann equation
becomes zero in the cubic root solutions. Indeed, the solutions exist for
arbitrary coupling ($g_{0}$), but zero potential ($V_{0}=0$).

Note that these are special solutions of the full system (they are singular
solutions). However what is important from the solution $a_{1}\left(
t\right)  $ is that power $p$ is not related with $\lambda\,$, as in the case
of a minimally coupled scalar field.\ However the solution of the latter is
recovered when we set $g_{0}=0$ and $p=\frac{2}{\lambda^{2}}$.

\subsection{Stability of the ideal gas solution}

Now, let's follow the approach of \cite{Liddle:1998xm,Uzan:1999ch} to
investigate the stability of the singular solution
\begin{align*}
a_{1}\left(  t\right)   &  =a_{0}t^{p}~,~\phi\left(  t\right)  =\frac
{2}{\lambda}\ln\left(  \phi_{0}t\right)  ~,~g_{0}=-\frac{\lambda\left(
\lambda^{2} p-2\right)  }{4 (3 p-1) \phi_{0}^{2}},~ V_{0} =\phi_{0}^{2}
\left(  \frac{2}{\lambda^{2}}+p (3 p-2)\right)
\end{align*}

Let's assume $\phi_{0}>0$. Introducing the new variables
\begin{equation}
\label{vars_scaling}\epsilon=\frac{\lambda\phi}{2\ln(t\phi_{0})}-1, \quad
v=t\left(  \frac{\lambda\dot{\phi}}{2\ln(t\phi_{0})}-\frac{\lambda\phi}%
{2t\ln^{2}(t\phi_{0})}\right)  ,\quad t\phi_{0}=e^{\tau},
\end{equation}
where $\phi(t)$ is a general solution of \eqref{f2}. That is, the scaling
solution $\phi_{s}\left(  t\right)  =\frac{2}{\lambda}\ln\left(  \phi
_{0}t\right)  $, corresponds to the critical point $\epsilon=0$. Furthermore,
by definition we have $v=\epsilon^{\prime}$, where now the prime denotes
derivative with respect to $\tau$. Using the variables \eqref{vars_scaling},
the equation \eqref{f2} recasts as the nonautonomous system
\begin{subequations}
\label{eq.41-42}%
\begin{align}
&  \epsilon^{\prime}=v,\\
&  v^{\prime}=-\frac{(\tau v+\epsilon+1)\left(  6(1-3p)p-\left(  \lambda
^{2}p-2\right)  e^{2\tau\epsilon}(\tau v+\epsilon+1)\left(  9p^{2}-3p-2(\tau
v+\epsilon+1)^{2}\right)  \right)  }{2\tau\left(  -\left(  \lambda
^{2}p-2\right)  e^{2\tau\epsilon}(\tau v+\epsilon+1)(3p-2(\tau v+\epsilon
+1))-3p+1\right)  }\nonumber\\
&  -\frac{(3p-1)\left(  \lambda^{2}p(3p-2)+2\right)  e^{4\tau-2\tau
(\epsilon+2)}}{2\tau\left(  -\left(  \lambda^{2}p-2\right)  e^{2\tau\epsilon
}(\tau v+\epsilon+1)(3p-2(\tau v+\epsilon+1))-3p+1\right)  }+\frac
{(\tau-2)v+\epsilon+1}{\tau}%
\end{align}
which admits the exact (singular) solution $(\epsilon,v)=(0,0)$.

Assuming $\epsilon\ll1,v\ll1$, we obtain the linearized nonautonomous system
\end{subequations}
\begin{equation}
\left(
\begin{array}
[c]{c}%
\epsilon^{\prime}\\
v^{\prime}%
\end{array}
\right)  =\left(
\begin{array}
[c]{cc}%
0 & 1\\
-\frac{(3p-1)(2\tau+1)}{\tau} & -3p-\frac{2}{\tau}+1
\end{array}
\right)  \left(
\begin{array}
[c]{c}%
\epsilon\\
v
\end{array}
\right)  . \label{eq.42}%
\end{equation}
The exact solution of \eqref{eq.42} is given by
\begin{subequations}
\label{Eq.44}%
\begin{align}
&  \epsilon(\tau)=e^{\frac{1}{2}\left(  1-3p+\sqrt{9p^{2}-30p+9}\right)  \tau
}\epsilon_{1}(\tau)+e^{\frac{1}{2}\left(  1-3p-\sqrt{9p^{2}-30p+9}\right)
\tau}\epsilon_{2}(\tau)\\
&  v(\tau)=e^{\frac{1}{2}\left(  1-3p+\sqrt{9p^{2}-30p+9}\right)  \tau}%
v_{1}(\tau)+e^{\frac{1}{2}\left(  1-3p-\sqrt{9p^{2}-30p+9}\right)  \tau}%
v_{2}(\tau).
\end{align}
where
\end{subequations}
\begin{align}
&  \epsilon_{1}(\tau)=\frac{c_{1}}{\tau},\;\epsilon_{2}(\tau)=\frac{c_{2}%
}{\tau},\\
&  v_{1}(\tau)=c_{2}\left(  \frac{1-3p+\sqrt{9p^{2}-30p+9}}{2\tau}-\frac
{1}{\tau^{2}}\right)  ,\;v_{2}(\tau)=c_{1}\left(  \frac{1-3p-\sqrt
{9p^{2}-30p+9}}{2\tau}-\frac{1}{\tau^{2}}\right)  .
\end{align}
The two modes of $\epsilon$ and $v$ are decaying for $p>1/3$, which are
exponentially depressed for $p>3$ and shows damped oscillations for $\frac
{1}{3}<p<3$. In both cases the perturbations decrease. For $p<1/3$ one mode is
decaying and the other grows as $\tau\rightarrow\infty$ (the origin is a
saddle), so that the perturbation increases with time.

For large $\tau$, the system \eqref{eq.42} can be approximated by
\begin{equation}
\label{eq.43}\left(
\begin{array}
[c]{c}%
\epsilon^{\prime}\\
v^{\prime}%
\end{array}
\right)  =\left(
\begin{array}
[c]{cc}%
0 & 1\\
-2 (3 p-1) & 1-3 p\\
&
\end{array}
\right)  \left(
\begin{array}
[c]{c}%
\epsilon\\
v
\end{array}
\right)  .
\end{equation}
The origin has the eigenvalues
\[
\left\{  \frac{1}{2} \left(  -\sqrt{9 p^{2}-30 p+9}-3 p+1\right)  ,\frac{1}{2}
\left(  \sqrt{9 p^{2}-30 p+9}-3 p+1\right)  \right\}  .
\]
The origin of \eqref{eq.43} is a stable node for $p\geq3$ or a stable spiral
for $\frac{1}{3}<p<3$, and this result seems to be independent on the
parameter $\lambda$.

\begin{figure}[ptb]
\centering
\includegraphics[width=0.7\textwidth]{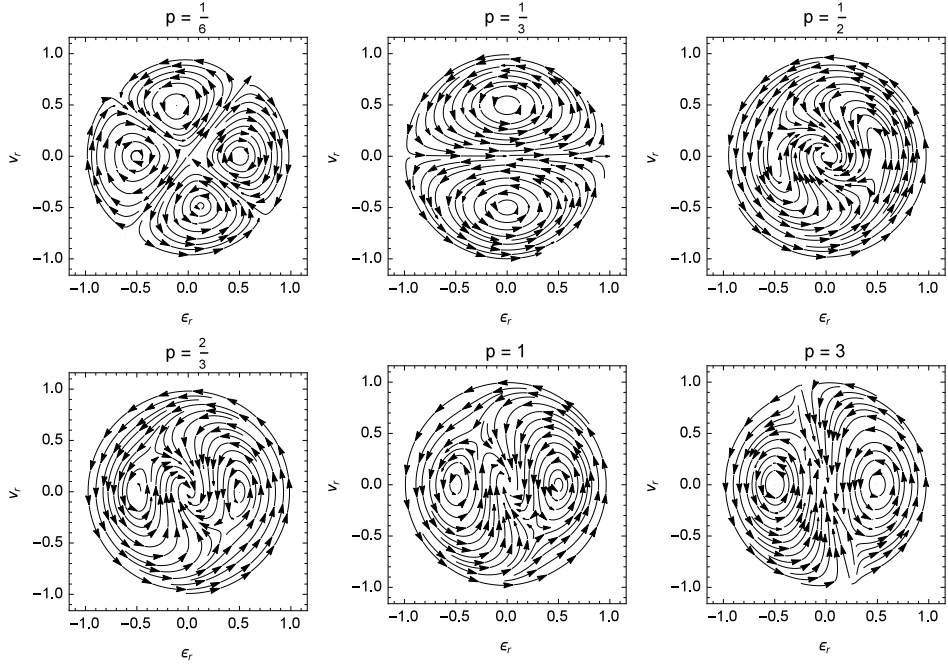} \caption{In the
figures are presented some orbits of \eqref{eq.43} on the Poincar\`{e} plane
for $p=1/6$, $p=1/3$, for radiation ($p=1/2$) and matter ($p=2/3$) dominated
universes, for zero acceleration solution ($p=1$) and for accelerated solution
($p=3$).}%
\label{fig:ProblemAFig1}%
\end{figure}

For the numerical integration we use the Poincar\`{e} projection:
\begin{equation}
\epsilon=\frac{r}{1-r}\sin\theta, \quad v =\frac{r}{1-r}\cos\theta,
\end{equation}
rescaling the time derivative by $f^{\prime}\rightarrow(1-2 r)^{2} r
f^{\prime}$ and plotting the solutions in the plane $(\epsilon_{r}%
,v_{r})=(r\sin\theta,r\cos\theta)$. This projection shrinks all the
trajectories in the phase plane to the unit disk. Furthermore, the points
$N\equiv(0,1)$, $E\equiv(-1,0)$, $W\equiv(1,0)$ correspond respectively to
$(\epsilon=0,\epsilon^{\prime}=\infty)$, $(\epsilon=-\infty,\epsilon^{\prime
}=0)$, and $(\epsilon=\infty,\epsilon^{\prime}=0)$. \newpage In the figure
\ref{fig:ProblemAFig1} it is presented some orbits of \eqref{eq.43} on the
Poincar\`{e} plane for different choices of $p$. In the first case, $p=1/6$,
the origin is a saddle and appear 4 non trivial centers. for $p=1/3$, the
origin is a saddle and the centers reduce to 2 and the line $v_{r}=0$ is
invariant. For $p=1/2$ the origin is the only stationary structure of the
phase space and it is a stable spiral. For $p=2/3,p=1$ the origin is a stable
spiral, but two centers appear, in addition to two saddle points. For $p=3$
the centers remain, but the origin now becomes a stable node. There appears
two saddle points at the interior of the phase space and two nonhyperbolic
points on the Poincar\`{e} circle which are saddle. 

Since the system \eqref{eq.43} is an approximation of \eqref{eq.42}, we may
expect that the origin of \eqref{eq.42} could be stable node for $p\geq3$ or a
stable spiral for $\frac{1}{3}<p<3$, and the result should be independent on
the parameter $\lambda$. Indeed, this result can be confirmed from the exact
solution, \eqref{Eq.44}, of the linear nonautonomous system. Then, since the
system \eqref{eq.42} is an approximation of \eqref{eq.41-42} for $\epsilon
\ll1, \epsilon^{\prime}\ll1$, we might expect analogous results for the full
system, at least near the origin. However, since the system \eqref{eq.41-42}
is non-autonomous and non-linear, this heuristic reasoning is not a complete
proof, and we have to test the validity of our results, which can be done
numerically by integrating the full system \eqref{eq.41-42}. Furthermore,
since the system is non-autonomous, the orbits cross in the Poincar\`e
representation, but not on the 3D representation of the integral curves
$(\epsilon, \epsilon^{\prime}, \tau)$.

Notice that the full system \eqref{eq.41-42} can be written as an autonomous
system by introducing the new variables
\begin{equation}
\mathcal{E}=e^{-\epsilon\tau}-1, \quad\mathcal{V}=\epsilon+\tau v,
\end{equation}
such that $(\epsilon,v)=(0,0)$ is mapped onto $(\mathcal{E},\mathcal{V}%
)=(0,0)$ for all $\tau$. And we define the new derivative $f^{\prime
}=e^{\epsilon\tau} \frac{d f}{d\tau}$. This leads to the dynamical system
\begin{subequations}
\label{eq.47}%
\begin{align}
&  \mathcal{E}^{\prime}=-\mathcal{V},\\
&  \mathcal{V}^{\prime}=\frac{\mathcal{V}^{2} \left(  \lambda^{2} p-2\right)
(2 \mathcal{V} (\mathcal{V}+2)-9 (p-1) p)}{2 (\mathcal{E}+1) \left(  3
\lambda^{2} p^{2} (\mathcal{V}+1)+3 p (\mathcal{E} (\mathcal{E}+2)-2
\mathcal{V}-1)-2 \lambda^{2} p (\mathcal{V}+1)^{2}+(-\mathcal{E}+2
\mathcal{V}+1) (\mathcal{E}+2 \mathcal{V}+3)\right)  }\nonumber\\
&  +\frac{(3 p-1) \left(  \mathcal{V} \left(  (\mathcal{E}-1) (\mathcal{E}%
+3)-3 \lambda^{2} p^{2}+p \left(  -3 \mathcal{E} (\mathcal{E}+2)+2 \lambda
^{2}+3\right)  \right)  +\mathcal{E} (\mathcal{E}+2) (\mathcal{E}
(\mathcal{E}+2)-3 p+3)\right)  }{(\mathcal{E}+1) \left(  3 \lambda^{2} p^{2}
(\mathcal{V}+1)+3 p (\mathcal{E} (\mathcal{E}+2)-2 \mathcal{V}-1)-2
\lambda^{2} p (\mathcal{V}+1)^{2}+(-\mathcal{E}+2 \mathcal{V}+1)
(\mathcal{E}+2 \mathcal{V}+3)\right)  }\nonumber\\
&  +\frac{\mathcal{E} (\mathcal{E}+2) (\mathcal{E} (\mathcal{E}+2)+2)
\lambda^{2} p (3 p-2) (3 p-1)}{2 (\mathcal{E}+1) \left(  3 \lambda^{2} p^{2}
(\mathcal{V}+1)+3 p (\mathcal{E} (\mathcal{E}+2)-2 \mathcal{V}-1)-2
\lambda^{2} p (\mathcal{V}+1)^{2}+(-\mathcal{E}+2 \mathcal{V}+1)
(\mathcal{E}+2 \mathcal{V}+3)\right)  }.
\end{align}

\begin{figure}[ptbh]
\centering
\includegraphics[width=0.60\textwidth]{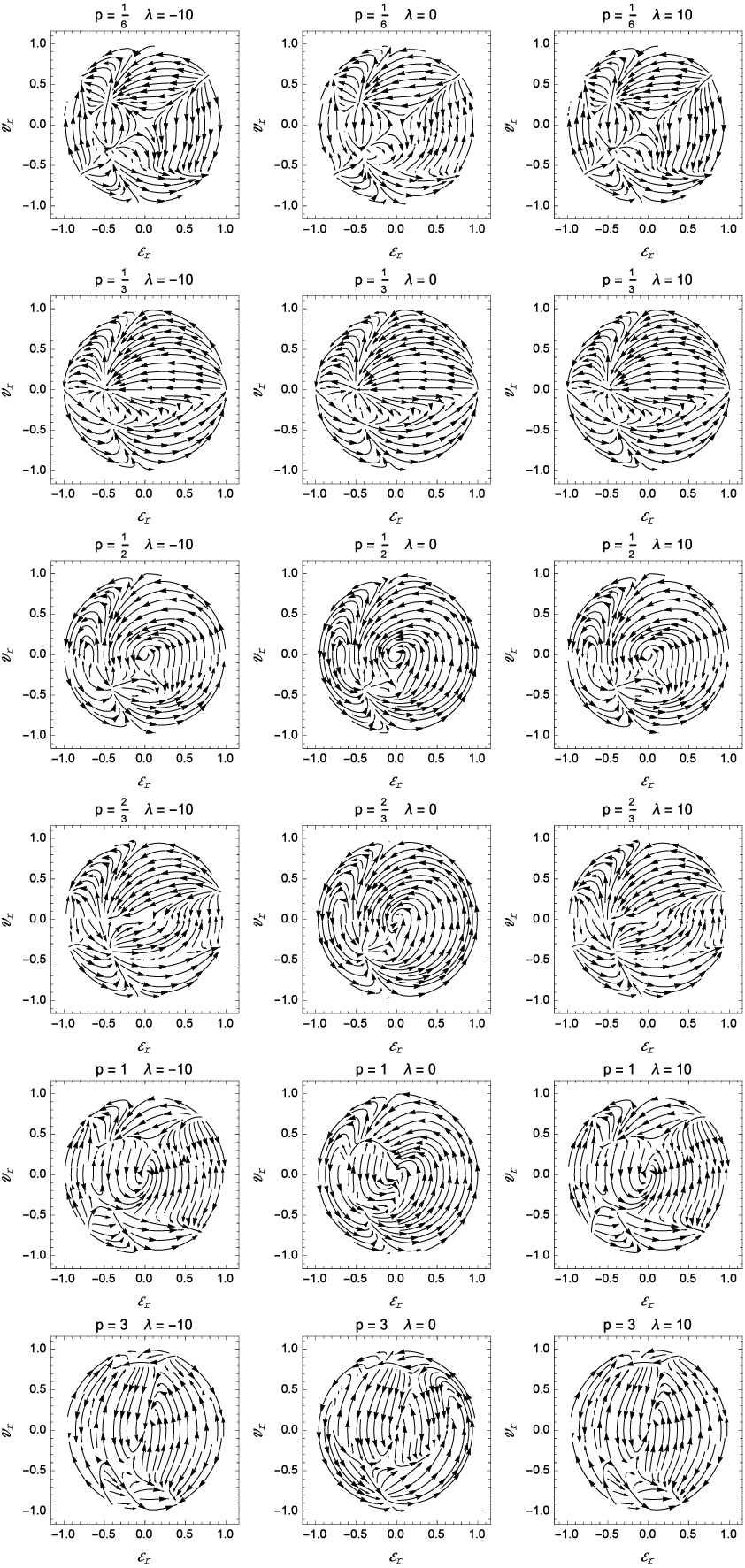}\caption{Plot of the
solutions in the plane $(\mathcal{E}_{r},\mathcal{V}_{r})=(r \sin\theta, r
\cos\theta)$ for the system \eqref{eq.47}.}%
\label{fig:ProblemAFig2}%
\end{figure}\newpage For the numerical integration we use the Poincar\`e
projection:
\end{subequations}
\begin{equation}
\mathcal{E}=\frac{r}{1-r}\sin\theta,\quad\mathcal{V}=\frac{r}{1-r}\cos\theta,
\end{equation}
and plot the solutions in the plane $(\mathcal{E}_{r},\mathcal{V}_{r})=(r
\sin\theta, r \cos\theta)$. This projection shrinks all the trajectories in
the phase plane to the unit disk. Furthermore, the points $N\equiv(0,1)$,
$E\equiv(-1,0)$, $W\equiv(1,0)$ correspond respectively to $(\mathcal{E}=0,
\mathcal{V}=\infty)$, $(\mathcal{E}=-\infty, \mathcal{V}=0)$, and
$(\mathcal{E}=\infty, \mathcal{V}=0)$.

In the figure \ref{fig:ProblemAFig2} are plotted some solutions in the plane
$(\mathcal{E}_{r},\mathcal{V}_{r})=(r \sin\theta, r \cos\theta)$ for the
system \eqref{eq.47}. These numerics support the claim that the phase spaces
are topologically equivalent for different choices of $\lambda$. Henceforth,
the stability results are independent of $\lambda$. For $p=1/2$ it is
confirmed that the origin is a saddle point as it is anticipated from the
analysis of the linearized non-autonomous system (and this result is valid for
all $p<1/3$). For $p=1/3$ the line $\mathcal{V}=0$ is invariant and it is not
stable. For the values $p=1/2, 2/3, 1$ the origin is a stable spiral as it is
anticipated from the analysis of the linearized non-autonomous system; these
results are true for $1/3<p<3$. Finally, for $p=3$ (and greater values of $p$)
the origin is a stable node. Comparing the results of figures
\ref{fig:ProblemAFig1} and \ref{fig:ProblemAFig2} we see that for the same
values of $p$ (and independently of $\lambda$), the dynamics near the origin
is topologically equivalent. However, the global features of the phase spaces
are rather different; for example, it seems from the diagrams that the system
\eqref{eq.47} has no center points as the system \eqref{eq.43}.

\subsection{Stability of the cubic root solution}

\begin{figure}[th]
\centering
\includegraphics[width=0.5\textwidth]{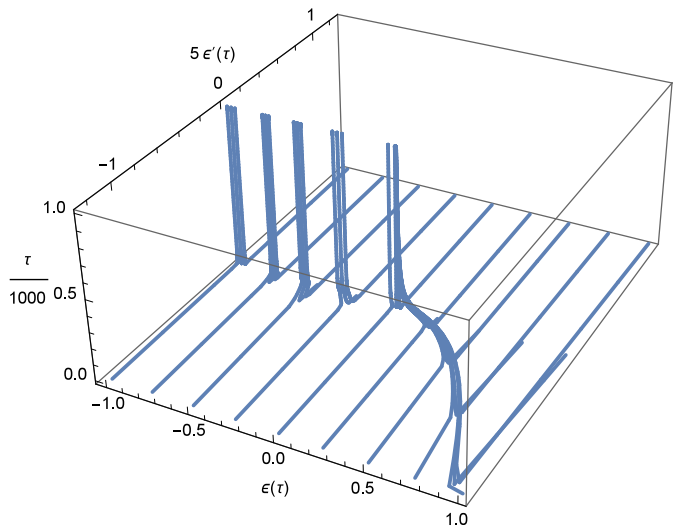} \caption{Some integral
curves for the full system \eqref{eq.41-42b}. It can be shown the $(0,0)$
solution is stable as $\tau\rightarrow\infty$ by choosing $c_{2}$ small enough
and for $\epsilon>0$.}%
\label{fig:ProblemBFig1}%
\end{figure}

\begin{figure}[ptbh]
\centering
\includegraphics[width=0.7\textwidth]{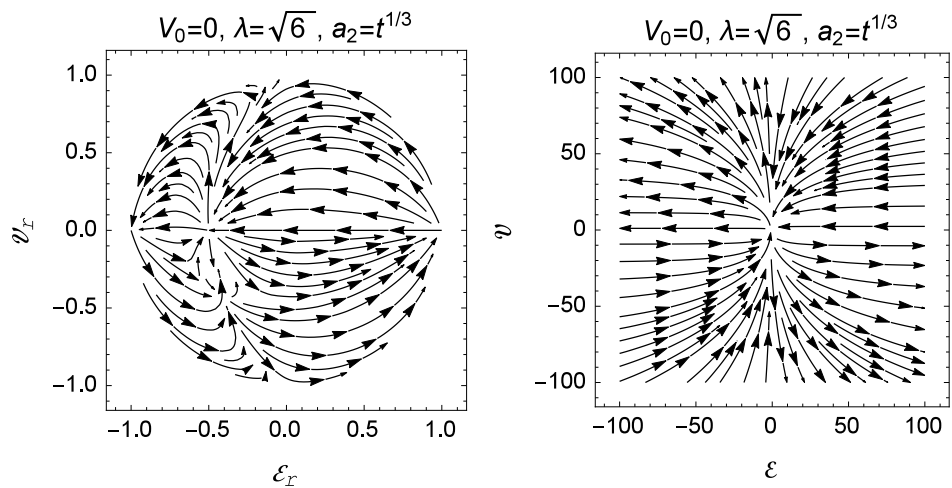} \caption{Poincar\`{e}
projection of the system \eqref{eq.55}. }%
\label{fig:ProblemBFig2}%
\end{figure}

For analyzing the stability of the cubic root solution (we choose just the
solution on the branch \textquotedblleft$+$\textquotedblright; the analysis
for $a_{3}(t)$ is quite similar):
\[
a_{2}\left(  t\right)  =a_{0}t^{\frac{1}{3}}~,~\phi\left(  t\right)
=\frac{\sqrt{6}}{3}\ln\phi_{0}t~,~V_{0}=0~,~\lambda=\sqrt{6},
\]
we substitute the values of $a_{2}(t)$, $V_{0}$ and $\lambda$ in \eqref{f2}
for an arbitrary $\phi$ to obtain
\[
3g_{0}e^{\sqrt{6}\phi}{\dot{\phi}}^{4}+{\ddot{\phi}}\left(  2\sqrt{6}%
g_{0}e^{\sqrt{6}\phi}{\dot{\phi}}^{2}-\frac{2g_{0}e^{\sqrt{6}\phi}{\dot{\phi}%
}}{t}+1\right)  +\frac{{\dot{\phi}}}{t}=0.
\]
Introducing the new variables
\begin{equation}
\label{vars_scaling-radiation_1}\epsilon=\frac{\sqrt{6}\phi}{2\ln(t\phi_{0}%
)}-1, \quad v=t\left(  \frac{\sqrt{6}\dot{\phi}}{2\ln(t\phi_{0})}-\frac
{\sqrt{6}\phi}{2t\ln^{2}(t\phi_{0})}\right)  , \quad t\phi_{0}=e^{\tau},
\end{equation}
where $\phi(t)$ is a general solution of \eqref{f1} and we have assumed
$\phi_{0}>0$. Without losing generality we can set $g_{0}\phi_{0}^{2}=1$,
which means that $g_{0}$ is given in units of $\phi_{0}^{-2}.$ Using the
variables \eqref{vars_scaling-radiation_1}, the equation \eqref{f1} recasts as
the nonautonomous system
\begin{subequations}
\label{eq.41-42b}%
\begin{align}
&  \epsilon^{\prime}=v,\\
&  v^{\prime}=\left(  1-\frac{2}{\tau}\right)  v-\frac{3(\tau v+\epsilon
+1)\left(  2\sqrt{\frac{2}{3}}e^{2\tau\epsilon}(\tau v+\epsilon+1)^{3}%
+1\right)  }{\tau\left(  2\sqrt{6}e^{2\tau\epsilon}(\tau v+\epsilon+1)(2\tau
v+2\epsilon+1)+3\right)  }+\frac{\epsilon+1}{\tau},
\end{align}
which admits the exact (singular) solution $(\epsilon,v)=(0,0)$. The
linearized equations are
\end{subequations}
\[
\epsilon^{\prime}=v,\;v^{\prime}=-\frac{2v}{\tau}.
\]
whose solution is
\[
\epsilon(\tau)=c_{2}-\frac{c_{1}}{\tau},v(\tau)=\frac{c_{1}}{\tau^{2}}%
\]
The solutions satisfy $\epsilon+v\tau=c_{2}$, where $c_{2}=\lim_{\tau
\rightarrow\infty}\epsilon(\tau)$. From these expressions we can have the the
$(0,0)$ solution is stable as $\tau\rightarrow\infty$ by choosing $c_{2}$
small enough. In the figure \ref{fig:ProblemBFig1} we present some integral
curves for the full system \eqref{eq.41-42} and the above feature is
illustrated. Now, from the full system \eqref{eq.41-42b}, we can obtain an
autonomous system by introducing the new variables
\begin{equation}
\mathcal{E}=e^{-\epsilon\tau}-1,\quad\mathcal{V}=\epsilon+\tau v,
\end{equation}
and the new derivative $f^{\prime}= e^{\epsilon\tau}\frac{df}{d\tau}$. This
leads to the dynamical system
\begin{subequations}
\label{eq.55}%
\begin{align}
&  \mathcal{E}^{\prime}=-\mathcal{V},\\
&  \mathcal{V}^{\prime}=-\frac{2\sqrt{6}\mathcal{V}^{2}(\mathcal{V}+1)^{2}%
}{3(\mathcal{E}+1)^{3}+2\sqrt{6}(\mathcal{E}+1)(\mathcal{V}+1)(2\mathcal{V}%
+1)}.
\end{align}
For the numerical integration we use the Poincar\`{e} projection:
\end{subequations}
\begin{align}
\mathcal{E}  &  =\frac{r}{1-r}\sin\theta,\nonumber\\
\mathcal{V}  &  =\frac{r}{1-r}\cos\theta,
\end{align}
and plot the solutions in the plane $(\mathcal{E}_{r},\mathcal{V}_{r}%
)=(r\sin\theta,r\cos\theta)$ as shown in figure \ref{fig:ProblemBFig2}.

\section{Conclusion}

\label{conc}

There are various ways to study the integrability of a dynamical system. In
this work we decided to search for conservation laws given by the application
of Noether's symmetries for the point-like Lagrangian of a Galileon field. For
that field we considered a model which can be seen an extension of the
minimally coupled canonical scalar field. Specifically the new term which \ is
introduced from the Galileon Lagrangian provides cubic powers of the
derivatives in the \textquotedblleft Klein-Gordon\textquotedblright\ equation
for the field, while a function is introduced such that, when it goes to zero,
the canonical scalar-tensor theories are recovered.

The unknown parameters of the model are two, the potential~$V\left(
\phi\right)  $ and the function $g\left(  \phi\right)  $ which is introduced
from the higher-power derivatives in the gravitational Lagrangian. These two
functions drive the evolution of the field equations and provide us with
different cosmological models. The demand that the field equations form an
integrable system with a conservation law linear in the momentum, is
sufficient to define the explicit form of these two unknown functions.
Specifically, the application of Noether's theorem implies that $V\left(
\phi\right)  $ and $g\left(  \phi\right)  $ are exponential such that
$V\left(  \phi\right)  \propto\left(  g\left(  \phi\right)  \right)  ^{-1}$.
This model is of special interest because, when the universe is dominated by
the scalar field potential, then the extra terms in the field equations which
corresponds to the Galileon field, do not affect the evolution of the system
and the Galileon field behaves like a canonical scalar field.

Due to the complexity of the field equations it was not possible to extract
the general solution in a closed-form. However, we proved the existence of
power-law solutions as special solutions. In order to study the evolution of
the system, we followed the method proposed by
\cite{Liddle:1998xm,Uzan:1999ch}. We found, after linearization, that for the
powelaw solution there are two modes of $\epsilon$ and $v$ which are decaying
for $p>1/3$, at an exponential rate for $p>3$; and the solutions manifest
damped oscillations for $\frac{1}{3}<p<3$. In both cases the perturbations
decrease. For $p<1/3$ one mode is decaying and the other grows as
$\tau\rightarrow\infty$ (the origin is a saddle), so that the perturbation
increases with time. We have constructed an autonomous dynamical system from
the full system. Using numerical integrations we showed that the stability
results are independent of $\lambda$; the dynamics near the origin is
topologically equivalent of that of the linearized system. However, the global
features of the phase spaces are rather different. For the cubic root solution
we found that the origin is stable but not asymptotically stable by choosing
proper initial conditions.

The cosmological eras of a canonical scalar field can be recovered from the
gravitation action (\ref{lan.03}) as it has been shown in \cite{genlyGL}.
However, even in that limit there are various differences between the two
models. One of the main that can be observed from the analysis which we
performed is in the relation of the exponent $p$, in the power-law solution
$a\left(  t\right)  =a_{0}t^{p}$, with the rate of the exponential potential.
Specifically for the canonical scalar field cosmology, the exponential
potential $V\left(  \phi\right)  =V_{0}e^{-\lambda\phi}$ admits a power-law
solution in which the power $p$ and the constant $\lambda$ are related as
follows $\lambda^{2}\simeq p^{-1}~$\cite{ellis1}. On the contrary, this is not
necessary for our model where we have shown that $p$ is independent of the
exponential rate of the potential; a fact that is owed to the cubic terms in
the Lagrangian.

It is true that power-law solutions describe those of an ideal gas but since
$p$ is now arbitrary the Galileon field can describe also fluids with a
negative equation of state parameter or even phantom fluids with equation of
state parameter smaller than minus one. Hence, from our analysis it follows
that the integrable model can describe the late-time acceleration of the
universe when the equation of state parameter for the dark energy is different
from minus one, or the early inflationary epoch demanding that the power-law
solution is unstable so that an exit from the inflationary period does exist.
That is in agreement with the various studies of Galileons, for instance see
\cite{inf1,infl2,infl3} and references therein. Finally, the stability
analysis of the power-law solutions differs from that of the canonical field
\cite{Liddle:1998xm} and that is directly related with the existence of the
qubic term in the Action Integral.

To show what is the role of the cubic term in the cosmological history let us
consider now that it dominates in the Lagrangian. Then, the field equations
reduce to the following relations
\begin{equation}
\ddot{\phi}+\frac{1}{2}\lambda\dot{\phi}^{2}\simeq0,\quad\dot{H}-\lambda
H\dot{\phi}+3H^{2}+\frac{1}{6}\lambda^{2}\dot{\phi}^{2}\simeq0
\end{equation}
with solution\footnote{Where we have chosen the initial conditions
$H(t_{0})=H_{0},\;\phi(t_{0})=\phi_{0},\;\dot{\phi}(t_{0})=\dot{\phi}_{0}%
$.\newline} $H(t)=\frac{6H_{0}-\lambda\dot{\phi}_{0}}{3\left(  (t-t_{0}%
)\left(  6H_{0}-\lambda\dot{\phi}_{0}\right)  +2\right)  }+\frac{\lambda
\dot{\phi}_{0}}{3\lambda\dot{\phi}_{0}(t-t_{0})+6},\quad\phi(t)=\phi_{0}%
+\frac{2}{\lambda}\ln\left(  1+\frac{\lambda\dot{\phi}_{0}}{2}(t-t_{0}%
)\right)  $.~Hence, for the scale factor follows
\begin{equation}
a(t)=a_{0}\left(  1+\frac{1}{2}\lambda\dot{\phi}_{0}(t-t_{0})\right)
^{\frac{1}{3}}\left(  1+\frac{1}{2}(t-t_{0})\left(  6H_{0}-\lambda\dot{\phi
}_{0}\right)  \right)  ^{\frac{1}{3}}.
\end{equation}
By taking initial conditions such that $\dot{\phi}_{0}=0$ or $6H_{0}%
-\lambda\dot{\phi}_{0}=0$, we obtain the cubic root solution that we derived
before $a(t)\propto t^{\frac{1}{3}}$. On the other hand, if we choose initial
conditions such that $3H_{0}-\lambda\dot{\phi}_{0}=0$, we obtain that
$a(t)\propto t^{\frac{2}{3}}$, that is, it corresponds to an universe
dominated by a dust fluid/dark matter.

In order to understand better the evolution and the dynamics of that model an
analysis of the critical points in the dimensionless variables should be
performed. With the latter we will able to investigate the effects of the
various terms of the Lagrangian in the evolution of the universe and to which
a eras they provided. The speciality of our model lies in the existence of the
second conservation law and it is of special interest to see how the second
conservation law fixes the evolution. On the other hand that special case is
not explicitly included in the results of \cite{genlyGL}. This is still work
in progress and will be published elsewhere.

Of course as we have mentioned before, only a special class of the generalized
Galileon models was studied in this work. Possible extensions by admitting
more terms in the action or more complicated coupling functions is a subject
of interest. The reason is that while some can numerically approximate the
evolution of the system; the existence of an actual solution of the field
equations is an open subject.

\begin{acknowledgments}
This work was financial supported by FONDECYT grants 3150016 (ND), 1150246
(AG) and 3160121 (AP). GL wants to thanks to DI-VRIEA by financial support
through Proyectos VRIEA Investigador Joven 2016 and Investigador Joven 2017.
AP thanks the Durban University of Technology and the University of the
Witwatersrand for the hospitality provided while this work was performed. SJ
would like to acknowledge the financial support from the National Research
Foundation of South Africa with grant number 99279. AP and SJ also acknowledge
support from the DST-NRF Centre of Excellence in Mathematical and Statistical
Sciences (CoE-MaSS). Opinions expressed and conclusions arrived at are those
of the authors and are not necessarily to be attributed to the CoE-MaSS.
\end{acknowledgments}

\appendix

\section{The Symmetry conditions}

\label{appenA}

In this Appendix for the convenience of the reader, we present the Noether
symmetry conditions (\ref{Lie.03}) for the Lagrangian (\ref{lag}) of the
Galileon model that we have considered.%

\[
\xi_{,a}=0,~\xi_{,\phi}=0,~\eta_{\phi,a}=0,~9g(\phi)a^{2}\eta_{\phi,t}=0,
\]%
\[
6\,{a}\eta_{a,t}-h_{,a}=0,~\,{a}^{3}\eta_{\phi,t}+f_{,\phi}=0,
\]%
\[
6\,a\eta_{a,\phi}-\,{a}^{3}\eta_{\phi,a}=0,
\]%
\[
3\,a^{2}g\eta_{a,t}-2\,g_{,\phi}{a}^{3}\eta_{\phi,t}=0,
\]%
\[
3\,\eta_{a}{a}^{2}V(\phi)+\eta_{\phi}{a}^{3}V_{,\phi}+{a}^{3}V\xi_{,t}%
-f_{,t}=0,
\]%
\[
-3a\xi_{,t}+3\eta_{a}+6a\eta_{a,a}=0,
\]%
\[
\frac{1}{2}a^{3}\xi_{,t}-\frac{3}{2}\eta_{a}a^{2}-a^{3}\eta_{\phi,\phi}=0,
\]%
\[
3g(\phi)a^{2}\eta_{a,\phi}-2g_{,\phi}a^{3}\eta_{\phi,\phi}+\frac{3}{2}%
g_{,\phi}a^{3}\xi_{,t}-\frac{3}{2}g_{,\phi}\eta_{a}a^{2}-\frac{1}{2}\eta
_{\phi}a^{3}g_{,\phi\phi}=0,
\]%
\[%
\begin{array}
[c]{c}%
6\eta_{a}g(\phi)a+3\eta_{\phi}a^{2}g_{,\phi}+3g(\phi)a^{2}\eta_{a,a}+\\
+9g(\phi)a^{2}\eta_{\phi,\phi}-2a^{3}g_{,\phi}\eta_{\phi,a}-9g(\phi)a^{2}%
\xi_{,t}=0.
\end{array}
\]

The solution of the system provides us with the symmetry vectors and the
specific forms of the functions $g\left(  \phi\right)  $ and $V\left(
\phi\right)  $ that were presented in section \ref{section3}.


\end{document}